\documentstyle[aps,twocolumn,epsfig]{revtex}

\title{Error free quantum communication through noisy channels}

\author{Anders S\o rensen and Klaus M\o lmer\\
{\small Institute of Physics and Astronomy, University of Aarhus}\\
{\small DK-8000 \AA rhus C}}

\begin{document}
\draft

\maketitle

\begin{abstract}
We suggest a method to  perform a quantum logic gate between distant
qubits  by  off-resonant field-atom dispersive interactions. The scheme we
present is shown to work ideally even in  the presence of errors in the
photon channels used for communication. The stability against errors arises
from the paradoxical situation that  the transmitted photons carry no
information about the  state of the
qubits.
In contrast to a  previous  proposal for ideal communication
[Phys. Rev. Lett. {\bf 78}, 4293 (1997)] our proposal only involves 
single atoms in the sending and receiving devices.  
\end{abstract}


\section{introduction}
Quantum mechanics is known to produce a variety of phenomena in lack of
classical  interpretation. In recent years the fields of
quantum computation and quantum communication have tried to exploit these
phenomena to propose computers and communication devices which are superior to
their classical
counterparts. One particular example is quantum teleportation which
is based on the nonlocal features of the EPR-paradox.
Quantum  teleportation is transmission of qubits
without actually sending the physical system, e.g. transfer of  the state 
of an atom to another atom at a different location. Classically
teleportation can be performed by
measuring the state of an object and sending the
information to the receiver who reconstructs the state in a similar
object. In the
quantum world it is not that easy. Quantum mechanics forbids us 
to gain exact knowledge of the state of an object. However, Bennett et al. have
suggested  that it is still possible to perform teleportation\cite{Bentele}, provided that
the transmitting system does not retain any information about the state
which is transmitted.
Recently quantum teleportation of a
photonic state has been achieved experimentally
\cite{expteleIns,expteleRom}. 

In practical realizations of teleportation the system may be subject to
noise in the transmission channels. Recently, van Enk et. al. have shown that the effect
of the  noise 
can be completely avoided, if we make suitable physical assumptions about the
noise in the channels \cite{enk}.

Eliminating noise on 
quantum information is considerably more complicated than eliminating noise on classical
information because quantum mechanics forbids copying of information 
$|\psi\rangle\rightarrow |\psi\rangle|\psi\rangle$, where $|\psi\rangle$ is
the state of a qubit ($c_0|0\rangle+c_1|1\rangle$)\cite{noklone}. However,
quantum mechanics does allow what we shall call a backup copy
$|\psi\rangle\rightarrow |\psi\rangle_a + |\psi\rangle_b$, where a single
quantum system is transfered to a state with projections on two different
subspaces $a$ and $b$, which are both equivalent to the initial state. (We
use  unnormalised
states except where otherwise stated). We call it backup copying because if
one half is ``lost'' (projected out), say the $b$ part, we may still have
the intact quantum
state in the $a$ part. (The exact  meaning of this statement will become
clear below). 

In this paper we use the backup encoding to perform quantum communication
in the presence of  errors in the channel used for
communication. Rather than considering teleportation as discussed by van Enk et. al., we 
perform a perfect control-not operation, which is slightly more general 
than teleportation.
To perform the operation we use off-resonant dispersive
interactions between atoms and the transmitted photons.  We assume that all errors are
due to imperfections in the transmission between atoms $\alpha$ and $\beta$ and
imperfections in the dispersive interaction, whereas
measurements and unitary evolutions on a single atom are assumed ideal. With this
assumption we show that our
scheme works
ideally  even in the presence of a
quite general class of errors.

We emphasize that our scheme is not a conventional quantum error
correcting code \cite{review}.  We use a specific physical model of the noise to remove
errors to all orders with a limited number of qubits whereas  conventional
error codes introduce new qubits
to correct errors up to a certain order.

\section{General description}

\begin{figure}
\centerline{
\epsfig{file=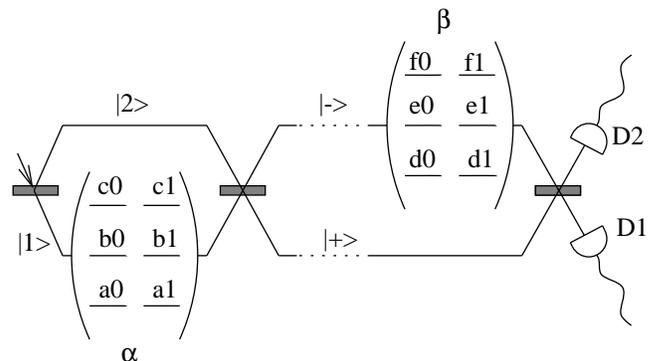,width=8.5 cm}}
\caption{Structure of the atoms $\alpha$ and $\beta$ and the suggested
  setup. The two atoms have three levels (denoted by letters) with a
  twofold Zeeman degeneracy (denoted by 0 and 1). The sending section
  consists of two beam splitters and the atom $\alpha$. The communication
  channels are the two dotted lines $|+\rangle$ and $|-\rangle$ and the
  receiving section is the atom $\beta$, the last beam splitter and the two
  detectors.}
\label{setup}\label{config}
\end{figure}

The control-not gate works between two atoms or ions. The control atom is 
called $\alpha$ and the target is called $\beta$. Here $\alpha$
and $\beta$ are two three level atoms, where each level has a two fold Zeeman
degeneracy ($J=1/2$).   The states of $\alpha$ 
are denoted by $ |a_i\rangle$, $|b_i\rangle$ and $ |c_i\rangle$ and the states
of $\beta$ are called $|d_i\rangle$, $|e_i\rangle$ 
and $|f_i\rangle$, where $i=0$ and 1 represents the azimuthal quantum numbers
$m=-1/2$ and $m=1/2$, see figure
\ref{config}. In practice, one may have recourse to systems with a
different arrangement of states, but our procedures are most easily
explained in the suggested realization.

If we consider the quantum information to be stored in the two Zeeman degenerate
ground state levels, the action of  the control-not operation
can be characterized by its action on suitable basis vectors for the atomic
ground state
\begin{equation}
  \label{cnot}
  \left[
    \begin{array}{cccc}
      |a_0\rangle |d_0\rangle\\
      |a_0\rangle |d_1\rangle\\
      |a_1\rangle |d_0\rangle\\
      |a_1\rangle |d_1\rangle
    \end{array}
  \right]
  \begin{array}{cccc}
    \rightarrow\\
    \rightarrow\\
    $\raisebox{-1ex}[0cm][0cm]{$\searrow \hspace{-1em} \nearrow$}$\\

   \end{array}
  \left[
    \begin{array}{cccc}
      |a_0\rangle |d_0\rangle\\
      |a_0\rangle |d_1\rangle\\
      |a_1\rangle |d_1\rangle\\
      |a_1\rangle |d_0\rangle
    \end{array}
  \right].
\end{equation} 
The control-not operation interchanges
the states $|d_0\rangle$ and $|d_1\rangle$  of $\beta$ if and only if
$\alpha$ is in the state  $|a_1\rangle$. A comment on notation:  Rather
than considering the evolution of a superposition of the four basis vectors
($c_{00} |a_0\rangle |d_0\rangle+c_{01}|a_0\rangle |d_1\rangle+
c_{10}|a_1\rangle |d_0\rangle+c_{11}|a_1\rangle |d_1\rangle$), we consider
the evolution of each basis vector. This emphasizes that each
vector in Eq. \ref{cnot} could be entangled with other qubits as in a
computational task. 

Our scheme consists of local encodings and two transmissions of photons from $\alpha$
to $\beta$. We begin with a local backup encoding on $\alpha$. We then
perform the first transmission followed by a symmetrization on
$\alpha$ and protection of relevant states of $\beta$. Another transmission
is performed  and finally we extract the
desired quantum states.

The effect of the transmissions is to  entangle the levels of the two atoms. By
performing  local operations we can then use this
entanglement to  implement the control-not operation. 

The stability of our
scheme arises from the horizontal symmetry among the atomic states in
figure \ref{config}. In the transmission we only use linearly ($\pi$) polarized
pulses which couple states vertically. This means that the photons contain no
information about whether $\alpha$ is in 0 or 1. The photons only contain
information about the levels of $\alpha$.
If, for instance, we start
with a superposition
$(c_0|a_0\rangle+c_1|a_1\rangle)+(c_0|b_0\rangle+c_1|b_1\rangle)$ and a photon
is absorbed during a transmission, the wavefunction will collapse to
some energy level (for instance $a$), but our quantum mechanical
superposition  between 0 and 1 will be
intact ($c_0|a_0\rangle +c_1|a_1\rangle$). From this ``backup'' state we
can start the transmission again and continue until we are successful.

We first describe the evolution in the ideal situation where we
have perfect error free transmissions.  

\section{Backup control-not under ideal conditions}
To perform the evolution described by Eq. \ref{cnot} we  suggest using  an
experimental setup as shown in figure  
\ref{setup}.  Our setup can
be divided into a sending section, a receiving section and the channels
 connecting them. 
The sending section consist of a
beam splitter, the atom $\alpha$ and another beam splitter.       The
receiving  part is the atom $\beta$,
a beam splitter and two photon detectors. All beam splitters are 50/50. The
channels are the two
photon lines connecting the two sections.
In a realistic implementation  it might be preferable to use a delay rather
than two distinct channels, but for the sake of clarity we apply two
lines in our analysis.

Initially the qubits are stored in the ground states $|a_i\rangle$ and
$|d_j\rangle$. We first
perform  a local backup encoding on $\alpha$.
With a linear $\pi /2$ pulse we take $|a_i \rangle$ to $|a_i \rangle +|b_i
\rangle$. 

We then perform the first photon transmission.  A single linearly polarized photon is split
into two  orthogonal states
$|1\rangle$ and $|2\rangle$ by a beam splitter. The field state $|1\rangle$
interacts non-resonantly  with the atom 
$\alpha$ coupling  the level $b$  
with a detuning $\Delta$ and a coupling constant g to a
higher lying state in the atom. The energy
shift of the level $b$ can be calculated in second order perturbation theory to be $\hbar
g^2/\Delta$. If we choose an interaction time $T=\Delta \pi /g^2$ the phase
of the state vector will change by $\pi$
if $\alpha$ is in the level  $b$. A phase change
of $\pi$ may not be realistic in an experiment, and  we shall relax this
assumption later.

We then recombine the two photon amplitudes  yielding the two states
$|+\rangle=|1\rangle+|2\rangle$ and  $|-\rangle=|1\rangle-|2\rangle$.

The receiving atom $\beta$ is prepared with    a
linearly  polarized $\pi /2$ pulse so that 
$|d_j\rangle$  is taken to $|d_j\rangle+|e_j\rangle$. The photon
state $|-\rangle$ now
couples $|e_j\rangle$ non-resonantly to a higher level  yielding a
conditional phase shift  of $\pi$ 
as described for atom $\alpha$, and we then 
apply a second  $\pi /2$ pulse so that if the field is in
the $|+\rangle$ state $\beta$
will be taken back to $|d_j\rangle$ by the last pulse, but if the field is in
the $|-\rangle$ state $\beta$ will be taken to $|e_j\rangle$ due to
the phase change induced by the field. Since the $|+\rangle$ and
$|-\rangle$ states correspond to $\alpha$ being in $a$ and
$b$ respectively, this will create the desired entanglement between $\alpha$
and $\beta$, but at this point the atoms are also entangled with the photon.

We get rid of the photon with
a  quantum eraser: The two photon 
states $|+\rangle$ and  $|-\rangle$ interfere yielding the two detector states $|D_1\rangle$ and
$|D_2\rangle$. We assume here that the  mirrors are aligned so that
$|D_1\rangle$ corresponds to the incoming state $|1\rangle$ \cite{alignmirror}. We then
perform  a measurement revealing whether the photon
 is in  $|D_1\rangle$ or $|D_2\rangle$. If $|D_1\rangle$ is measured we change the sign
 of the level $b$ to compensate a sign induced by the eraser.

A simple analysis shows that the transmission performs the operation
\begin{eqnarray}
    |a_i \rangle |d_j \rangle \rightarrow |a_i \rangle |d_j \rangle
    \nonumber \\
    |b_i \rangle |d_j \rangle \rightarrow |b_i \rangle |e_j \rangle.
  \label{transmission}
\end{eqnarray}
During transmission
the Zeeman degeneracy plays no
role. Subscripts i and j denoting the Zeeman state have  only been written for
later  convenience.

Had we included the evolution of the level $e$, the transmission would be a
control-not  between  the levels $a$,$b$,$d$ and $e$, 
but this
control- not will be vulnerable to errors. Our
backup scheme makes it possible to perform a perfect control-not
between the states $|a_0\rangle$, $|a_1\rangle$, $|d_0\rangle$ and
$|d_1\rangle$, also  in the presence of
errors. Paradoxically this may be achieved by means of the  transmission described by
Eq. \ref{transmission} and local operations, even though the transmitted
photons carry no information on the azimuthal quantum numbers. 

Including the $\pi/2$ preparation of $\alpha$ the evolution so far is given by
\begin{equation}
  \left[
  \begin{array}{cccc}
    |a_0 \rangle |d_0 \rangle\\
    |a_0 \rangle |d_1 \rangle\\
    |a_1 \rangle |d_0 \rangle\\
    |a_1 \rangle |d_1 \rangle
  \end{array}
  \right]
  \rightarrow
  \left[
  \begin{array}{cccc}
    |a_0 \rangle |d_0 \rangle+ |b_0 \rangle |e_0 \rangle \\
    |a_0 \rangle |d_1 \rangle+ |b_0 \rangle |e_1 \rangle\\
    |a_1 \rangle |d_0 \rangle+ |b_1 \rangle |e_0 \rangle \\
    |a_1 \rangle |d_1 \rangle+ |b_1 \rangle |e_1 \rangle
  \end{array}
  \label{backupintermidiate}
  \right].
\end{equation}

Now, the states $|e_i \rangle$ are moved to  storage states
$|f_i \rangle$ and the states  $|a_i \rangle$ and $|b_i \rangle$ are
interchanged by linearly polarized $\pi$-pulses.  A second  photon is
transmitted, causing again the evolution in Eq. \ref{transmission}. Since
the  $f$-states of $\beta$ are not coupled to the incident photon, these
states are not affected by the second transmission, and we end up with 
\begin{equation}
  \left[
  \begin{array}{cccc}
    |b_0 \rangle |e_0 \rangle+ |a_0 \rangle |f_0 \rangle \\
    |b_0 \rangle |e_1 \rangle+ |a_0 \rangle |f_1 \rangle\\
    |b_1 \rangle |e_0 \rangle+ |a_1 \rangle |f_0 \rangle \\
    |b_1 \rangle |e_1 \rangle+ |a_1 \rangle |f_1 \rangle
  \end{array}
  \label{extract}
  \right] .
\end{equation} 
The main result of this paper is that
we are  able to construct these states
even in the presence of errors. This will be shown in sec. \ref{noisycnot}.
Within the  quantum states in Eq. \ref{extract} we  break the
horizontal symmetry of azimuthal states 0 and 1 and extract the
desired   states on
the right side of 
Eq. \ref{cnot} by local operations.

We measure if $\alpha$ is in the subspace spanned by
$|a_0\rangle$ and $|b_1\rangle $. This can for instance be done by
interchanging $|a_1\rangle$ and $|b_1\rangle$ and making a QND measurement \cite{qnd}
of the  energy of $\alpha$. If $\alpha$ is found in
the  subspace spanned by
$|a_0\rangle$ and $|b_1\rangle $ we
interchange the amplitudes on $|e_0\rangle$ and $|e_1\rangle$. If it is not we interchange
the amplitudes on $|f_0\rangle$ and $|f_1\rangle$.
We then measure  if
$\beta$ is in the  subspace spanned by  $|e_0\rangle + |f_0\rangle$ and
$|e_1\rangle  + |f_1\rangle$. This can be done
with a $\pi/2$ pulse followed by a QND measurement of the atomic energy. From the
results of these measurements one can construct a sequence of pulses
which takes us to the desired states.

As a specific example of the extraction procedure consider the situation
where  $\alpha$ is found in the 
subspace spanned by $|a_0\rangle$ and $|b_1\rangle$. The measurement
collapses Eq. \ref{extract} to this subspace and we apply a pulse which
interchanges $|e_0\rangle$ and $|e_1\rangle$ 
\begin{equation}
  \left[
  \begin{array}{cccc}
    |a_0 \rangle |f_0 \rangle\\
    |a_0 \rangle |f_1 \rangle\\
    |b_1 \rangle |e_0 \rangle\\
    |b_1 \rangle |e_1 \rangle
  \end{array}
  \right]
  \begin{array}{cccc}
    \rightarrow\\
    \rightarrow\\
    $\raisebox{-1ex}[0cm][0cm]{$\searrow \hspace{-1em} \nearrow$}$\\ 

  \end{array}
  \left[
  \begin{array}{cccc}
    |a_0 \rangle |f_0 \rangle\\
    |a_0 \rangle |f_1 \rangle\\
    |b_1 \rangle |e_1 \rangle\\
    |b_1 \rangle |e_0 \rangle
  \end{array}
  \right] .
  \label{extractex}
\end{equation} 
Now consider the situation where we measure that $\beta$ is in the
subspace spanned by $|e_0\rangle-|f_0\rangle$ and
$|e_1\rangle-|f_1\rangle$. Since
$|e_i\rangle$ can be written $(|e_i\rangle+|f_i\rangle)+(|e_i\rangle-|f_i\rangle)$ and
$|f_i\rangle$ can be written $(|e_i\rangle+|f_i\rangle)-(|e_i\rangle-|f_i\rangle)$ this is
seen to introduce a minus on the first two lines in Eq. \ref{extractex}. By
subsequently
transferring $|b_1\rangle$ to $-|a_1\rangle$ and $|e_i\rangle-|f_i\rangle$ to
$|d_i\rangle$ we arrive in the desired states.

This extraction procedure is  illustrated in figure
\ref{extfig}. We recall that the qubits are represented as superpositions of the
azimuthal quantum states $0$ and $1$. Entanglement between the atoms is
visualized by shading: Part (a) of  the figure illustrates the states
in Eq. \ref{extract}, where $\beta$ is in the level $e$ ($f$) if
$\alpha$ is in $b$ ($a$). Our first measurement chooses states
in $\alpha$ diagonally (part (b) in the figure and Eq. \ref{extractex}). Now, we interchange
azimuthal  states $0$ and $1$ of
$\beta$ if $\alpha$ is in $1$. From the shading in the figure this is seen
to correspond to interchanging $|e_0\rangle$ and $|e_1\rangle$
($|f_0\rangle$ and $|f_1\rangle$ if the other diagonal had been measured). 
Finally, all states are taken to the lowest level as described in the
example after Eq. \ref{extractex} and we end
up in the desired states. 

\section{Error analysis}
\label{error}
In this section we analyse the effect of errors and  in the next section we
show how our backup scheme eliminates these errors.
We will assume that measurements and  unitary evolutions in  single atoms
are perfect.  All errors will be 
due to imperfections in the dispersive interactions and in the channels used
for communication.

\subsection{Errors due to loss of photons}
A photon is considered lost if it is not detected at the photon detectors
in the end. Since the photons carry no information on the azimuthal quantum number the 
superposition between $0$ and $1$ states will not be disturbed. Using a QND measurement it
can be detected whether $\alpha$ is in $a$ or $b$ and the
qubit will still be present horizontally. Similarly we can measure the energy
of $\beta$ without disturbing the  qubit and the initial states can  be
restored.  We  can then start over again and proceed with
the transmission   until it is successful.

\subsection{Errors without loss of photons}
\noindent {\it Phase shift in the dispersive interaction with $\alpha.$} We
no longer  assume that our dispersive interaction causes  a phase shift which is $\pi$. With a
general phase shift the two levels $a$ and $b$ no longer give two
orthogonal photon states ($|+\rangle$ and $|-\rangle$) which can be
separated by a beam splitter. But we can arrange our beam splitter so that
$a$ always produces a photon in the $|+\rangle$ channel. The atom in level $b$,
however, will yield a superposition of $|+\rangle$ and $|-\rangle$
\begin{eqnarray}
    |a_i\rangle &\rightarrow & |a_i\rangle |+\rangle  \nonumber \\
    |b_i\rangle &\rightarrow & |b_i\rangle (|-\rangle +k_+|+\rangle).
 \label{errordisp}
\end{eqnarray}
We show in section \ref{noisycnot} that our scheme still works because the
erroneous $k_+ |+\rangle$ component can be projected out with a measurement.

\noindent {\it Errors in the channels.} 
With the assumption that the photons cannot jump from one channel to the
other and cannot be created in the channels, the most general  evolution
will  be described by 
\begin{eqnarray}
     |+\rangle &\rightarrow & \eta \widetilde {|+\rangle} \nonumber \\
    |-\rangle &\rightarrow & \zeta \widetilde {|-\rangle},
 \label{channels}
\end{eqnarray}
where the notation means that the wavepacket is changed in some way
(change of shape and duration of the
wavepacket
etc.). The
states $\widetilde {|+\rangle}$ and $\widetilde {|-\rangle}$ are 
assumed normalized. This evolution can be described by the non hermitian
Hamiltonian of Monte Carlo wavefunctions \cite{montecarlo} in the no jump
stages of evolution.

We assume
that the photon in two subsequent transmissions couples to independent and
identical environments. With this assumption the
evolution in Eq. \ref{channels} will be  the same in the two
transmissions 
provided  that the photon is not lost. This
assumption is further justified in \cite{enk}.

\noindent {\it Errors in the interaction with $\beta$.} The photon state $\widetilde{|+\rangle}$
does not interact with the atom $\beta$ and  we assume that local laser pulses on
$\beta$ are error free. This photon state will therefore not cause any
transition in $\beta$. The interaction between $\beta$ and $\widetilde{|-\rangle}$
is modified due to the imperfect dispersive
interaction and the modified
photon state. 
This means that $\beta$ may not be transfered to the level
$e$ as desired. The effect of the  interaction may be summarized
as follows
\begin{eqnarray}
  \label{fejlram}
    \widetilde{|+\rangle}|d_i\rangle & \rightarrow &
    \widetilde{|+\rangle}|d_i\rangle \nonumber \\
    \widetilde{|-\rangle}|d_i\rangle & \rightarrow & 
    \widetilde{|-\rangle}  (k_d|d_i\rangle +|e_i\rangle).
\end{eqnarray}

\noindent {\it Errors in the photon detection.} The two orthogonal states
$\widetilde{|-\rangle}$  and
$\widetilde{|+\rangle}$  are measured in
an orthogonal basis \{$|D_1\rangle$, $|D_2\rangle$\}. The beam splitter is a 50-50
beam splitter and any overall phase factors may be absorbed in the definition of
$|D_1\rangle$ and $|D_2\rangle$. We can therefore write
\begin{eqnarray}
  \label{fejlbeamsplit}
     |D_1\rangle&=&\widetilde{|+\rangle}+e^{i\delta}\widetilde{|-\rangle}
     \nonumber \\
    |D_2\rangle&=&\widetilde{|+\rangle}-e^{i\delta}\widetilde{|-\rangle}.
\end{eqnarray}
The phase factor $\delta$ in the two equations must be identical because
$|D_1\rangle$ and $|D_2\rangle$ have to be orthogonal.

Collecting the effects of Eqs. \ref{errordisp} through
\ref{fejlbeamsplit} we see that (before photon detection) the transmission
performs  the evolution 
\begin{eqnarray}
   |a_i\rangle|d_j\rangle & \rightarrow& \eta  |a_i\rangle|d_j\rangle
   (|D_1\rangle+|D_2\rangle) \nonumber \\
   |b_i\rangle|d_j\rangle &\rightarrow&  \zeta
   e^{-i\delta}|b_i\rangle|e_j\rangle(|D_2\rangle-|D_1\rangle)+|b_i\rangle|d_j\rangle \nonumber\\
  && \times [\zeta k_d
   e^{-i\delta}(|D_2\rangle-|D_1\rangle)+\eta k_+ (|D_1\rangle+|D_2\rangle)].
 \label{errortrans}
\end{eqnarray}
This expression displays unwanted disturbances of the amplitudes of our
quantum mechanical superposition. As we shall see below, these disturbances
can be interchanged in the second photon transmission, thereby symmetrizing and hence
eliminating their effect on the relevant amplitudes. 

\section{Noisy backup control-not}
\label{noisycnot}
We now  describe the effect of errors on the overall evolution.
If a photon is lost we restore the initial situation and start the
transmission again as described above.  
In this section we shall therefore only consider the situation where we do not loose
photons. 

After the first
transmission we end up in states like Eq. \ref{errortrans}. We recall that we
change the sign of the level $b$  if $D_1$ clicks and after photon detection
 the atomic state will therefore be given by
\begin{equation}
\left[
\begin{array}{cccc}
|b_0\rangle [\zeta e^{-i\delta}(|e_0\rangle+k_d |d_0\rangle)\pm \eta k_+ |d_0\rangle]\\
|b_0\rangle [\zeta e^{-i\delta}(|e_1\rangle+k_d |d_1\rangle)\pm \eta k_+ |d_1\rangle]\\
|b_1\rangle [\zeta e^{-i\delta}(|e_0\rangle+k_d |d_0\rangle)\pm \eta k_+ |d_0\rangle]\\
|b_1\rangle [\zeta e^{-i\delta}(|e_1\rangle+k_d |d_1\rangle)\pm \eta k_+ |d_1\rangle]
\end{array}
\right]
+
\eta \left[
\begin{array}{cccc}
|a_0\rangle |d_0\rangle \\
|a_0\rangle |d_1\rangle\\
|a_1\rangle |d_0\rangle\\
|a_1\rangle |d_1\rangle
\end{array}
\right]
,
\end{equation}
where the sign on the $k_+ |d_i\rangle$ component is $+$ ($-$) if  $D_2$ 
($D_1$) clicks. 
We now interchange $|a_i\rangle$ and $|b_i\rangle$ and $|e_i\rangle$
and $|f_i\rangle$  before we
 perform the second transmission and subsequent photon detection. The  atomic states
will now read
\begin{eqnarray} 
\left[
\begin{array}{cccc }
|a_0\rangle [\zeta\eta e^{-i\delta} (|f_0\rangle+ k_d |d_0\rangle) \pm
\eta^2 k_+ |d_0\rangle]\\
|a_0\rangle [\zeta\eta e^{-i\delta} (|f_1\rangle+ k_d |d_1\rangle) \pm \eta^2 k_+ |d_1\rangle]\\
|a_1\rangle [\zeta\eta e^{-i\delta} (|f_0\rangle+ k_d |d_0\rangle) \pm \eta^2 k_+ |d_0\rangle]\\
|a_1\rangle [\zeta\eta e^{-i\delta} (|f_1\rangle+ k_d |d_1\rangle) \pm \eta^2 k_+ |d_1\rangle]
\end{array}
\right] \nonumber
\\
+
\left[
\begin{array}{cc}
|b_0\rangle [\eta\zeta e^{-i\delta}(|e_0\rangle+k_d |d_0\rangle) \pm \eta^2 k_+ |d_0\rangle]\\
|b_0\rangle [\eta\zeta e^{-i\delta}(|e_1\rangle+k_d |d_1\rangle) \pm \eta^2 k_+ |d_1\rangle]\\
|b_1\rangle [\eta\zeta e^{-i\delta}(|e_0\rangle+k_d |d_0\rangle) \pm \eta^2 k_+ |d_0\rangle]\\
|b_1\rangle [\eta\zeta e^{-i\delta}(|e_1\rangle+k_d |d_1\rangle) \pm \eta^2 k_+ |d_1\rangle]
\end{array}
\right],
\label{fejlslut}
\end{eqnarray}
where the $\pm$ in the first (second) square bracket refers to the outcome
of the photon detection in the first (second) transmission.
Eq. \ref{fejlslut} shows that we have achieved the desired symmetrization
of amplitude errors. Collecting terms  we get states of
the form $\eta\zeta
e^{-i\delta}(|a_i\rangle|f_j\rangle+|b_i\rangle|e_j\rangle)+|R_i\rangle|d_j\rangle$,
where $|R_i\rangle|d_j\rangle$ are all the remaining components. The first term is the ideal
states in Eq. \ref{extract}. However, we also have the $|R_i\rangle|d_j\rangle$
component. 

We now measure if $\beta$ is in the level $d$. If $\beta$ is found in $d$, the qubits are restored to their
initial states and the transmission is attempted again. If
$\beta$ is not in $d$, the $|R_i\rangle|d_j\rangle$ components are projected out by the
measurement and we are  left with the
states of Eq. \ref{extract}. 
From here the ``diagonal'' extraction proceeds as before.

\section{Discussion}
Above we have shown how to achieve a perfect quantum control-not operation through
noisy channels. It has been shown \cite{gates} that any  unitary
operation on any number of qubits 
can be performed using single qubit operations and control-not operations.  
With a perfect control-not we are therefore able to perform  any
communication task.

As mentioned in the introduction, van Enk et. al. have used similar
ideas to achieve perfect teleportation \cite{enk}. However, we believe that we suggest
a simpler physical realization.
The coherent control of
several atoms required in the scheme of van Enk et. al. is a very difficult experimental
task, and it is a major advantage of our scheme that it only requires single atoms at each
end. 

In \cite{vanenkgate} van Enk et. al. also discus the possibility to
make an error free quantum logic gate using only single atoms. The main
idea in \cite{vanenkgate} is to monitor the performance of the gate and
discard unsuccesfull operations. Failures are monitored by a third state of
the atoms which, however, does not enable the recovery of  the quantum
information as in this  work and in \cite{enk}.

To perform our one atom scheme we have chosen to use non-resonant
dispersive interactions.  It is also possible to use other
kinds of physical
interactions, like Raman pulses as in the  suggestions
of van Enk et. al. Our only requirement is that 
that the
states $|d_i\rangle$ and $ |e_i\rangle$ in $\beta$ are coupled only when
$\alpha$ is in level $b$.

\begin{figure}
\centerline{
\epsfig{file=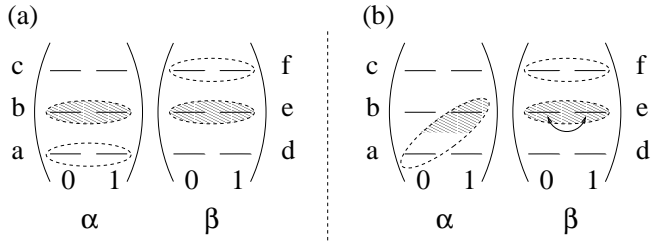,angle=270,width=8.5cm}}
\caption{Illustration of the diagonal extraction step. Part (a) corresponds
  to Eq. \ref{extract}. The qubits are present horizontally in the Zeeman
  states and the levels are entangled vertically (represented by the
  shading). In part (b) we make a measurement which chooses states of
  $\alpha$ diagonally. With the measurement outcome in the figure, the
  control-not is achieved by interchanging $|e_0\rangle$ and $|e_1\rangle$
  and transfering the atoms to the lower levels $a$ and $d$ as described in
  the text.}
\label{extfig}
\end{figure}

We wish to emphasize another important feature of our proposal, well illustrated in
figure \ref{extfig}. The use of atoms with two plus two relevant states,
rather than pairs of atoms with two times two states,
offers a simple geometric picture of the transfer protocol,  cf. in
particular the diagonal extraction in figure \ref{extfig} (b).
We believe that such pictures may be useful in the development of further
ideas, not only for fault-tolerant transmission.

As an example, consider computation distributed on several quantum
computers \cite{distributed}, with signalling atoms responsible for
communication. Following our proposal these atoms may be entangled
vertically, prior to the calculation, and when ready for transmission,
horizontal qubits may be communicated by the diagonal extraction procedure and 
other local operations.

Also multi-particle entanglement may be accommodated following these lines.
Recently it has been shown that for quantum communication over long
distances the efficiency of a channel can be enhanced if it
consists of series of nodes which  share EPR-pairs
with each of their neighbouring nodes \cite{repeaters}.  
To share EPR-pairs with two neighbours would normally require two
atoms per node. However, with our scheme  a single atom may suffice. If we start
with a superposition $|a_0\rangle-|a_1\rangle$ and perform a horizontal
control-not with one neighbour, these two nodes will share a horizontal EPR-pair. By
performing the steps which lead to Eq. \ref{extract} with another neighbour
a vertical EPR correlation with this neighbour is created without
destroying the horizontal correlation with the first neighbour. In this way
each node  only requires a single atom.

This work was completed under the newly established Thomas B. Thriges Center
for Kvanteinformatik at the Institute of Physics and Astronomy and the
Institute of Computer Science,
University of Aarhus.

\begin{thebibliography}{95}


\bibitem{Bentele} C. H. Bennett, G. Brassard, C. Cr\'epeau, R. Jozsa, A. Peres
  and W. K. Wooters, Phys. Rev. Lett. {\bf 70}, 1895 (1990).
 

\bibitem{expteleIns} D. Bouwmeester, Jian-Wei Pan, K.
  Mattle, M. Eibl, H.  Weinfurter and A. Zeilinger, Nature, {\bf
    390} (1997), 575.

\bibitem{expteleRom} D. Boschi, S. Branca, F. De Martini, L. Hardy and
  S. Popescu, Phys. Rev. Lett. {\bf 80}, 1121 (1998).

\bibitem{enk} S. J. van Enk, J.I. Cirac and P. Zoller, Phys. Rev. Lett. {\bf 78}, 4293 (1997).

\bibitem{noklone} W. K. Wootters and W. H. Zurek, Nature, {\bf 299}, 802 (1982).

\bibitem{review} See for instance A. Steane, Report No. quant-ph/9708022.

\bibitem{alignmirror} This assumption is only made for mathematical
  convenience. Our backup encoding is able to
  correct any phase errors corresponding to an arbitrary alignment of the
  mirror.

\bibitem{qnd} For an introduction to QND measurements see V. B. Braginsky, Y.
  I. Vorontsov and K. S. Thorne, Science, {\bf 209}, 547 (1980). In our
  setup a QND measurement of the energy of $\alpha$ could be achieved by
  injecting new photons and measuring whether the photons appear in the
  $|+\rangle$ or $|-\rangle$ channel.

\bibitem{montecarlo} K. M\o lmer and Y. Castin, Quantum Semiclass. Opt.,
  {\bf 8}, 49 (1996). 

\bibitem{gates} A. Barenco, C. H. Bennett, R. Cleve, D. P. Divincenzo,
  N. Margolus, P. Shor, T. Sleator, J. A. Smolin and H. Weinfurter,
  Phys. Rev. A {\bf 55}, 3457 (1995).

\bibitem{vanenkgate} S. J. van Enk, J. I. Cirac and P. Zoller,
  Phys. Rev. Lett. {\bf 79}, 5178 (1997)

\bibitem{distributed} A. K. Ekert, S. F. Huelga, C. Macchiavello and J
  .I. Cirac, Report No. 9803017;
  H. Buhrman, R. Cleve and W. van Dam, Report No. quant-ph/9705033 

\bibitem{repeaters} H.-J. Briegel, W. D\"ur, J.I. Cirac and P. Zoller,
  Report No. quant-ph/9803056.

\end{thebibliography}
\end{document}